\documentclass[aps,pra,onecolumn,superscriptaddress,longbibliography,nofootinbib,12pt]{revtex4-2}
\usepackage[]{amsmath}
\usepackage{graphics}
\usepackage{amssymb}
\usepackage{graphicx,epstopdf}
\usepackage[usenames,dvipsnames,svgnames]{xcolor}
\usepackage{appendix}

\newcommand {\ket}[1]{\lvert \, #1\rangle}

\newcommand {\braket}[2]{\langle #1 \, | \, #2 \rangle}

\begin{document}

\title{Stimulated absorption of single gravitons: first light on quantum gravity \vspace{10pt}}
\author{Victoria Shenderov}
\affiliation{Department of Physics, Stevens Institute of Technology, Hoboken, NJ 07030, USA}
\affiliation{Cornell University, Ithaca, NY 14853, USA}
\author{Mark  Kanex}
\affiliation{Department of Physics, Stevens Institute of Technology, Hoboken, NJ 07030, USA}
\affiliation{Massachusetts Institute of Technology, Cambridge, MA 02139, USA}
\author{Thomas Beitel}
\affiliation{Department of Physics, Stevens Institute of Technology, Hoboken, NJ 07030, USA}
\author{Germain Tobar}
\affiliation{Department of Physics, Stockholm University, SE-106 91 Stockholm, Sweden}
\author{Sreenath K. Manikandan}
\affiliation{Nordita, KTH Royal Institute of Technology and Stockholm University, SE-106 91 Stockholm, Sweden}
\author{Igor Pikovski\vspace{0.5cm}}
\email{Corresponding author: pikovski@stevens.edu}
\affiliation{Department of Physics, Stevens Institute of Technology, Hoboken, NJ 07030, USA}
\affiliation{Department of Physics, Stockholm University, SE-106 91 Stockholm, Sweden} \vspace{0.5cm}

\begin{abstract}
\vspace{.5cm}
    In a recent work we showed that the detection of the exchange of a \textit{single} graviton between a massive quantum resonator and a gravitational wave can be achieved. Key to this ability are the experimental progress in preparing and measuring massive resonators in the quantum regime, and the correlation with independent detections of gravitational waves (GWs) that induce stimulated absorption, 
    such as by the Laser Interferometer Gravitational Wave Observatory (LIGO). But can  stimulated single-graviton processes shed light on the quantization of gravity? Here we analyze this question and make a historic analogy to the early days of quantum theory. We discuss in what ways such experiments can indeed probe key features of the quantized interaction between gravity and matter. We outline five experimental tests utilizing stimulated emission and absorption processes of single gravitons, 
    opening the first window into experimental exploration of the fundamentals of  quantum gravity.   
\end{abstract}

\maketitle

\section{Introduction} 
\begin{figure}[t]
    \includegraphics[width=0.9\linewidth]{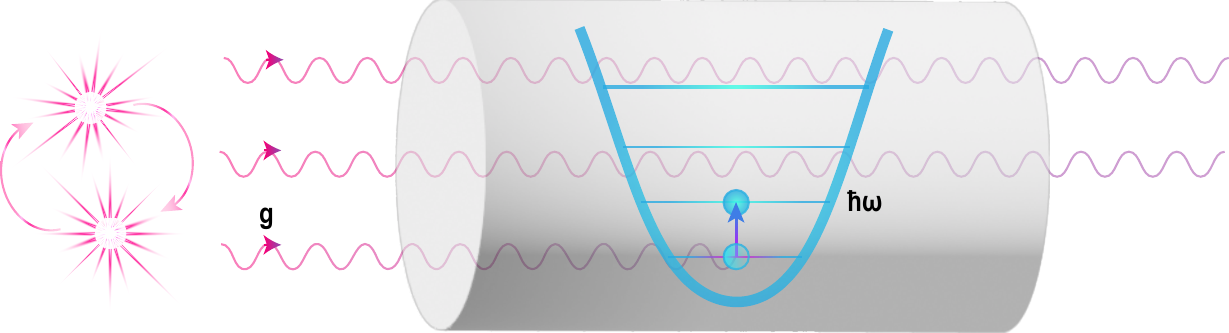}
    \caption{ 
    A single-graviton-detector consists of a massive system with quantized energy levels that can be resolved, and prepared in an initial energy eigenstate such as the ground state. The weak interaction with a passing gravitational wave can result in resonant absorption of a single graviton of energy $E=\hbar \nu$ matching the detector energy $E=\hbar \omega$. The correlation to independent detection of the passing GW indicates the stimulated single graviton process. This detector principle was outlined in Ref. \cite{tobar2023detecting} and may be feasible in near-future experiments.  }
    \label{fig:detector}
\end{figure}

The quantization of gravity implies the existence of a fundamental particle associated with the gravitational interaction --- the graviton. Even though no full theory of quantum gravity has yet been found, the prediction of a graviton is a core feature in most approaches. Based on regular quantized linear gravity, a graviton is a massless boson that mediates the gravitational interaction, and which carries spin-2 and no other charge. In practice, a graviton also constitutes an indivisible quantum of energy $E=hf$ that forms gravitational waves of frequency $f$, and where $h$ is Planck's constant. While gravitational waves have now been directly confirmed with LIGO from numerous compact binary mergers \cite{abbott2016observation, abbott2017gw170817}, such waves are highly energetic and deeply in the classical regime. Nevertheless, it was recently shown by some of us that the detection of a single graviton is possible and realistic with near-future quantum technology \cite{tobar2023detecting}. One key element is that due to the very small scattering cross-section, of order $\sigma \sim l_p^2$ \cite{Weinberg1972} for single atoms, where $l_p \approx 10^{-35}$~m is the Planck length, only very few gravitons from such a classical wave actually interact with the detector. It is therefore possible to design a resonant macroscopic system that would absorb only a \textit{single} graviton from a passing gravitational wave --- but which is still laboratory-scale. Detecting this single event also requires the ability to prepare the quantum ground state of the massive resonator, time-continuous sensing of quantum jumps in discrete energy steps in the detector, and the cross-correlation to LIGO events to confirm absorption from the independently measured gravitational wave (it turns out neutron-star-mergers \cite{abbott2017gw170817} are the best candidates), as illustrated in Fig. \ref{fig:detector}. While challenging, these capabilities are realistically achievable, and thus single-graviton detection can be realized with kg-scale resonators \cite{tobar2023detecting}, potentially augmented with lower-mass transducers for quantum sensing \cite{tobar2024detecting}.

Here we discuss how such graviton detection capabilities can offer first empirical input on the fundamentals of linearized quantum gravity. In essence, such a detection would confirm the stimulated absorption or emission of a single graviton. But how much can we learn about quantum gravity from such a stimulated event, extracted from a very energetic classical wave? It is helpful to draw an analogy to the early tests of quantum theory at the turn of the 20th century. Graviton detection would be analogous to the photoelectric effect, where stimulated absorption of photons was observed \cite{lenard1902ueber}, and which is the basis of modern-day photon-detectors. In 1905, Einstein proposed the quantization of light and applied this hypothesis to explain the photoelectric effect \cite{einstein1905erzeugung}, which sparked the development of quantum theory. After much resistance, the quantum of light was fully accepted in the mid 1920ies. Today, however, it is well-known that stimulated emission such as in the photoelectric effect can be described with semi-classical optics \cite{lamb1968photoelectric}. Stimulated single-quantum transitions alone do not conclusively prove that the radiation is a quantum field.  Remarkably, a wide range of quantum phenomena can be explained through semi-classical models, in which matter is quantized but radiation is treated classically, including spontaneous emission, the Lamb shift, and vacuum polarization \cite{mandel1976ii} (but with the drawback that conservation laws have to be abandoned).  A rigid, modern benchmark for the existence of photons is thus the measurement of quantum statistics of light, which can indicate the quantum nature of the radiation field through sub-Poissonian statistics or a vanishing intensity-intensity correlation function $g^{(2)}(0)$, because of the indivisibility of a single photon energy \cite{clauser1974experimental}. An even more stringent benchmark is the characterization of the quantum state of the incoming radiation, with quantum features typically attributed to a negative Wigner function rather than a negative Glauber-Sudarshan-P-function that is sufficient for non-classical statistics \cite{vanner2015towards}, or even the violation of Bell's inequalities as the ultimate test of non-classicality \cite{aspect1982experimental}. Given these modern developments, it has been a natural question whether tests of quantum gravity can demonstrate such benchmarks\footnote{After the first version of this manuscript, it was shown that any potential quantum statistical features of the radiation are directly accessible in this type of detector \cite{sreenath2025,manikandan2025complementary,toccacelo2026quantum}.}. 
Realizing some of the most stringent tests available to quantum optics today would be very challenging for the graviton case \cite{carney2024graviton}.

Here, instead, we argue that a more modest approach that relies only on stimulated processes can still experimentally reveal a great deal about the possible quantization of gravity. We first briefly review our results \cite{tobar2023detecting} that show how single gravitons can be detected through stimulated absorption, and then discuss how this can be used to infer and test aspects of quantum gravity. We draw a historic analogy to the early days of quantum mechanics \cite{pais1979einstein,jammer1989conceptual,mehra2000historical}, when the existence of the photon was well established and accepted by the physics community long before the development of full QED, let alone modern quantum optical tests. 
We then propose tests of fundamental principles of single-graviton exchange, discussing what aspects of quantum gravity can be probed with such capabilities and similar experiments. Our results show that important foundational questions about the quantized interaction between matter and gravity can be probed, offering critical experimental input on quantum gravity. 

\section{Single-graviton detection}

Whether a graviton can be detected even in principle has been a fascinating question for a long time. The conventional answer has been that this is nearly impossible. Weinberg \cite{Weinberg1972} (and later Boughn and Rothman \cite{BoughnStephen2006Aogd}) for example, showed that atomic transitions involving gravitons would proceed roughly only once every $10^{40}$~s, while Dyson argued that LIGO would need a 37 orders of magnitude improvement in sensitivity to detect single gravitons (Dyson estimates the number of gravitons in a typical gravitational wave by dividing the wave's energy density $\rho_E= \frac{c^2}{32 \pi G} h^2 \nu^2$ by the energy density of a single graviton with $E= \hbar \nu$ in a cubic box of size $c/\nu$) \cite{dyson2013graviton}, both vastly impractical. We revisited this question \cite{tobar2023detecting}, and as summarized below, we show that inferring energy exchanges at the level of a single graviton is much more practical than previously thought. Before outlining this result, we wish to briefly comment on why neither LIGO nor individual atoms are good probes for inferring energy exchanges at the level of a single graviton. Firstly, LIGO measures displacements with very high precision, but this is not the best quantum mechanical observable to address the question of energy quantization. To infer energy exchanges in discrete amounts, one should think of a sensor that is sensitive to discrete changes in energy eigenstates, rather than continuous position. Secondly, atomic systems have traditionally been used to test quantum mechanics, but today quantum experiments are no longer confined to the atomic realm. For the interaction with gravity, the mass quadrupole is the relevant charge, thus scaling up to massive systems can be of great advantage. Finally, it is not necessary to improve the sensitivity to correspond to a single graviton strain amplitude for inferring the existence of gravitons. While low intensity measurements are crucial for many modern table top experiments and quantum information tasks, most quantum experiments in fact include bright sources. A historically important example is the photoelectric effect in which bright light shining on metal surfaces excites electrons, from which Einstein inferred the existence of a light-quantum. Dyson in fact also considered stimulated absorption for gravitons \cite{dyson2013graviton}, but in his scenarios it remained impractical. Instead, in Ref. \cite{tobar2023detecting} we looked at several modifications to previous consideration, which enable single graviton detection. One key element is the use of a collective mode in a massive quantum system, as opposed to a single atom. Today such massive systems can be brought into quantum states, and quantum mechanical effects are explored on macro-scales \cite{bose2023massive}. That would amount to the use of a quantum version of classic Weber-bar detectors. But it is also necessary to combine it with appropriate quantum sensing capabilities to monitor discrete changes in energy, just as for an atom \cite{haroche2006exploring}, as opposed to the conventional transduced position measurements. Notably there are still Weber bars in operation which look for classical gravitational wave signals, mostly at high frequencies \cite{AggarwalNancy2021Caoo}, but traditional designs can operate below kHz-frequencies \cite{aguiar2011past}. Finally, we circumvent the limitation of not being able to create gravitational waves by correlating the signal to independent LIGO detections: this provides heralded detection with independent confirmation of the amplitude and frequency of the incoming gravitational wave. However, because occurrence of such events is a priori unknown, a continuous quantum sensing scheme is required \cite{jordan2024quantum}.

The starting point of our analysis is deriving the interaction Hamitonian for a 1D collective quantum mode interacting with the gravitational wave, which is \cite{tobar2023detecting}
\begin{equation} \label{eq:Hamiltonian}
H_{int} = -\frac{1}{2} h_{\mu \nu} T^{\mu \nu} =  M L \ddot{h}(t) \sum_{l=1,3,...} \frac{\left( -1\right)^{(l+1)/2}}{\pi^2 l^2} \zeta_l - \frac{M \ddot{h}(t)}{8} \sum_{l} \zeta_l^2 \, .
\end{equation}
Here $\zeta_l$ are the collective acoustic modes (labeled by $l$), and $h(t)$ is the generally chirping gravitational wave. The quadrupole moments of each atom are oscillations about their mean position, which add up to produce the collective coupling.  The leading contribution to the interaction increases with the total mass $M$ of the system and its length $L$. The collective enhancement of each acoustic mode dominates over the second term, the tidal force on each atom by the gravitational wave as considered by Weinberg and others \cite{Weinberg1972,BoughnStephen2006Aogd,FischerUwe1994Tpfa}. It becomes a sub-leading correction.

Using this interaction Hamiltonian one can compute the absorption and emission processes.  While spontaneous emission remains untenable, one can achieve stimulated emission rates as good as 1 Hz: For example with a $1800$~kg Al-cylinder, stimulated by a gravitational wave with amplitude $h = 5 \times 10^{-22}$. Thus in each detection event, only a few gravitons are exchanged. Importantly, we can solve the dynamics exactly, and find that an initial ground state of the massive system evolves to $ \ket{0} \rightarrow \ket{\beta(t) e^{-i \omega t}}$ with the coherent state amplitude
\begin{equation} \label{eq:beta}
|\beta(t)| = \sqrt{\frac{M L^2}{\pi^4 \omega \hbar}} \, \left|  \int_0^t ds \ddot{h}(s) e^{-i \omega s} \right| = \sqrt{\frac{M L^2}{\pi^4 \omega \hbar}} \, \, \chi(h(t), \omega, t) \, .
\end{equation}
From this, we can deduce the probability of detecting an energy exchange at the level of a \textit{single} graviton when measured in the energy (Fock) basis:
\begin{equation}\label{eq:prob}
P_{0 \rightarrow 1} = \left| \braket{1}{\beta(t) e^{-i \omega t}} \right|^2 = \left| \beta(t) \right|^2 e^{- \left| \beta(t) \right|^2}.
\end{equation}
The maximum probability of a single graviton absorption is thus $P_{0 \rightarrow 1}^{(max)} = 1/e \approx 36\%$ for $\left| \beta(t) \right| = 1$. From this we obtain the optimal mass required for a bar detector to infer a single graviton exchange event, which expressed in terms of the speed of sound $v_s$ is given by
\begin{equation} \label{eq:mass}
M_{opt} = \frac{\hbar \pi^2 \omega^3}{v_s^2 \chi^2(h(t), \omega, t)} \, .
\end{equation}
The exact mass depends on the gravitational wave profile through $\chi$ as given in eq. \eqref{eq:beta}, which can either be found from available data, or by an analytic approximation for binary mergers \cite{tobar2023detecting}. For sources like the GW170817 neutron star --- neutron star merger as detected by LIGO \cite{abbott2017gw170817}, this yields an ideal mass on the order of $M_{opt} \approx 20$~kg for beryllium composition. Thus ground-state-cooled kg-scale detectors can be used to observe the absorption of a single graviton, if their energy levels are monitored through continuous quantum measurement. Such monitoring could also be implemented at much lower masses through transducers \cite{tobar2024detecting}. 

We note that noise supersession is also key, but it is feasible given the very short relevant window of cross-correlation to the classical detection events. For example, the NS-NS mergers detected at LIGO \cite{abbott2017gw170817} last only a few seconds within the detection band, and thus false positive counts in the click detector have to be suppressed only in this time window. This can be achieved with a mechanical $Q$-factor of roughly $Q=10^{10}$ and an environment temperature of $T=1$~mK, as discussed in Ref. \cite{tobar2023detecting}. In contrast, the mechanical Q-factor does not pose a limitation for the detector response to the signal itself. This is because the dependence on the quality factor typically arises for the steady-state between incident wave and detector absorption --- but for the aforementioned binary star mergers, this regime is never reached. Instead, the short time of the GW duration within the resonance window is of relevance (see also Appendix A), which enters in the optimal mass in eq. \eqref{eq:mass} through the GW-dependent function $\chi$.

As a simple example, for a monochromatic incident wave at angular frequency detuned from the detector's resonant frequency by $\Delta$, and for $\Delta \ll \omega$, the excitation probability becomes $P_{0 \rightarrow 1} = \frac{ h^2 M L^2 \omega^3 t^2}{4 \hbar \pi^4} \textrm{sinc}^2(t \Delta /2)$. In the long-time limit and on resonance, this yields the usual Fermi's Golden rule. This expression highlights the direct correspondence to the main features of the photoelectric effect: The probability of excitation, or the number of absorption events, increases with the GW intensity $h^2$, but the energy of transition is completely determined by the resonance condition $\Delta \approx 0$. The energy changes in discrete amounts at or near the resonant frequency. This is the well-known dependence on frequency and independence on intensity from the photoelectric effect. What is missing is the ability to measure the matter excitations through the voltage --- instead one can monitor the quantum jumps directly through time-continuous quantum non-demolition measurements of energy \cite{jordan2024quantum} and thus deduce the exchange of a single graviton. While challenging, energy measurements in massive quantum systems, albeit at lower masses, have been demonstrated \cite{von2022parity}.

Thus in summary, a single graviton detection is possible in the not-so-distant future, given the rapid advancements in quantum acoustics. But apart from improved technology requirements (quantum sensing in low-noise massive resonators and ground state cooling), the detection principle relies on stimulated emission from incoming gravitational waves. It thus cannot serve as a proof of the quantization of gravity --- stimulated processes can be described in the semi-classical limit. This is analogous to the Rabi-limit of the interaction Hamiltonian between light and matter: the gravitational field in eq. \eqref{eq:Hamiltonian} can be treated as a classical parameter, while the matter is quantized. In this limit, the same excitation probabilities are recovered. To go beyond the limit, it would be necessary to probe the quantum states of the radiation field or other quantum features, as discussed in the introduction. 

Nevertheless even stimulated processes can provide indirect evidence of the quantum nature of gravity, just as for the quantum nature of light in the first half of the 20th century. Clearly the exchange of quanta implies the change of energy in discrete amounts from the gravitational wave, thus assuming energy conservation to hold for each individual event, one can deduce that indeed gravitons are absorbed. But it is also useful to revisit the historic arguments for the acceptance of quantum theory. This analogy is very instructive, as the experiments on gravity we envision today are in many ways technologically constrained to indirect hints of quantumness as the early tests of quantum phenomena. Thus, rather than using modern benchmarks for non-classicality, here we focus on the historic arguments and experiments around quanta of light. This is not to suggest that the exact same reasoning would hold for gravity --- but rather to revisit what level of experimental evidence was attainable to make the existence of photons likely, long before modern quantum optical tests.
A direct link to quantum gravity can then be made. Our work showed that the exchange of quanta between matter and gravitational waves can be observed, but a probe of modern, stringent benchmarks of quantum behavior of gravity seems out of reach. Revisiting the historic arguments that showed early on the existence of a photon 
with high confidence can thus serve as a blueprint to explore questions about quantum gravity with similarly limited experimental capabilities in the near future.

\section{Historic perspective}

Historically, Planck's black-body calculation in 1900 showed the discrete exchange of energy between light and matter, in packages of $E=h f$ \cite{planck1901law}. Nevertheless, it was only Einstein that took the leap in 1905 to postulate that light actually consists of such discrete energy quanta, and that it is not simply limited to an anomalous interaction. Einstein realized that Planck's law required the modification of equipartition as predicted by classical electromagnetic theory, which would have led to what is today known as the Rayleigh-Jeans black-body behavior. Instead, he focused on the experimentally verified Wien's distribution law for high frequencies of the black-body spectrum. He realized that this law -- the high-frequency limit of Planck's distribution -- can be derived as if radiation behaved thermodynamically as ideal point particles, each with energy $E=hf$. This yields the correct change in entropy as function of volume, in direct analogy to an ideal Boltzmann gas, thus suggesting a gas of particles of light. He thus made the leap from the hypothesis of a quantized interaction between light and matter, to postulating that light itself behaves as if consisting of quanta \cite{einstein1905erzeugung, einstein1906theorie}. Einstein applied his hypothesis of quantized interaction to the photoelectric effect. Nevertheless, despite the success of his predictions, the theory of quantized light was not accepted for close to 20 years. Even after his famous measurements of Planck's constant through Einstein's formula for the photoelectric effect in 1916 \cite{millikan1916direct}, Millikan still considered Einstein's quantized radiation hypothesis ``untenable'' and even ``reckless'' \cite{pais1979einstein}. Thus, while the quantized exchange of energy was gradually accepted, the leap to a quantization of the \textit{free} radiation field was a much bigger challenge.

What then, beyond the photoelectric effect but long before quantum optical tests, led to the acceptance of quantized radiation? And what lessons can we draw for the gravitational case, for which some of the modern quantum optical tests are far out of reach? The developments after 1905 largely focused on the quantization of matter and the behavior of solids: Einstein realized that his quantum hypothesis, if applied to matter alone just as for light, implies a specific heat of materials that reproduced experiments at the time, but gave new predictions at low temperatures \cite{einstein1907plancksche}. Such experiments started to be realized in the early 1910s and confirmed Einstein's predictions, which drew attention to his quantum theory. Much of the work then focused on developing a more detailed quantum theory for matter, most notably Bohr's atomic model. Nevertheless, a quantum theory for radiation seemed unnecessary to most, including Planck, Millikan and even Bohr himself. Even Einstein had his doubts, but by 1917 was fully convinced about the reality of light quanta. Two developments were key in the final acceptance of quantized radiation.

Firstly, Einstein realized in 1917 that the quantized momentum also plays a key role \cite{einstein1917quantentheorie}. Light must also carry the discrete momentum $p=hf/c$, in addition to discrete energy $E=hf$. He realized this property when considering energy fluctuations in the context of molecular equilibrium under radiation pressure, in connection also with his famous result on the spontaneous and stimulated emission and absorption coefficients $A_{21}$, $B_{21}$, and $B_{12}$. The direct experimental confirmation of quantized momentum came through the observation of the Compton effect \cite{compton1923quantum}, which is inconsistent with scattering of a classical wave. 

Secondly, alternative views have become increasingly nonviable. Early attempts to explain the photoelectric effect without photons, such as Lenard's triggering hypothesis \cite{lenard1902ueber}, were dismissed due to experimental contradictions to such models. It was realized by Einstein and others, that if giving up energy conservation, one can explain the photoelectric effect statistically without light quanta but that this is undesirable\footnote{At the first 1911 Solvay conference, Einstein remarked: ``One can choose between the [quantum] structure of radiation
and the negation of an absolute validity of the energy conservation law. [...] We will agree that the energy principle
should be retained.'' \cite{einstein1911theorie}}. This in fact remains a feature of the modern quantum optical semi-classical limit, where energy is not conserved for individual absorption or emission events \cite{mandel1976ii}. In our proposal to detect single gravitons \cite{tobar2023detecting}, we argue precisely on these grounds of energy conservation that single graviton exchange can be inferred, when single quantum jumps are recorded. However, it is possible to envision energy-non-conserving alternatives. An early one in the context of electromagnetic radiation was put forward by Bohr, Kramers and Slater (BKS) in 1924 \cite{bohr1924lxxvi}, as a means to avoid the need for photons. It attempted to keep quantized matter without quantizing the field, conserving energy and momentum \textit{on average}, but violating these conservation laws for the individual exchange of quanta between field and matter. It is in this spirit similar to today's semi-classical limit. But this theory was quickly dismissed experimentally through Bothe's and Geiger's coincidence measurements between scattered photons and electrons, showing event-by-event conservation of energy and momentum \cite{bothe1925wesen}.
 
Thus, from limited but diverse experimental input and without direct observation, it was possible to confirm the existence of quanta of light. Some lingering doubts remained, which today we might call ``loopholes''. For example by Erwin Schr\"{o}dinger \cite{schrodinger1924bohrs,schrodinger1927comptoneffekt}, who considered the possibility of a classical electromagnetic wave which would Bragg-scatter from matter-waves to explain the Compton effect, and keeping the spirit of BKS theory by abandoning conservation laws for single events. This eventually motivated modern quantum optical coincidence experiments demonstrating sub-Poissonian statistics \cite{clauser1974experimental,short1983observation}, anti-bunching \cite{kimble1977photon} and Hong--Ou--Mandel interference \cite{hong1987measurement}, putting most reservations to rest. Nevertheless, it should be stressed that long before tests of quantum statistics or QED, quantization of radiation was confirmed to a high degree. While semi-classical models were able to account for some experimental features, they generally failed to capture different observations \cite{bosanac1998semiclassical}. This is the benchmark we wish to apply to the gravitational case today. 

\section{Tests with single-graviton interactions}

From the above historical developments and considerations, one can draw inspiration to design tests of quantum theory, even without the firm proof that modern quantum optical tests would offer. There is much to be explored even in the linearized limit of quantum gravity, and the single-graviton detection offers direct input on the regime of physics where gravity interacts with matter at the level of individual quanta. As briefly mentioned above, the expected quantization of Einstein's gravity makes firm predictions on the properties of these particles: it should be a single, massless gravitational particle, propagating at speed $c$, have spin-2, and carry energy $E=hf$. For each of these particle predictions, there is a corresponding collective behavior that explains the classical domain, and so there are indirect tests and constraints on these properties from current GW observations and consistency with classical gravity \cite{abbott2021tests}. However, this quantum-classical correspondence only works if one assumes the standard quantum framework. A \textit{direct} test of these properties of a single graviton is conceptually different: it would give input on the standard quantum predictions and its alternatives directly at the level of individual quanta. Importantly, there are many different theories on the nature of quantum gravity, some with their own concrete predictions on the nature of gravitons (however, many approaches are still mathematical and do not give concrete physical predictions). For example, several theories on symmetry breaking with Nambu-Goldstone modes predict a finite number of massive gravitons in addition to a massless one \cite{arkani2003effective,bluhm2005spontaneous}. Many bi-metric theories predict two spin-2 gravitons, one massive and one massless \cite{max2017gravitational,gialamas2023bimetric}. $f(R)$ models, in certain cases, predict a second particle called a scalaron to account for an additional scalar degree of freedom found in the linearized representation \cite{casado2025propagating}. And entropic gravitational theories lack gravitons altogether, instead predicting that gravity is an emergent phenomenon without quantization \cite{verlinde2011origin,carney2025quantum}. There are countless other possibilities for properties of gravitons, once the standard linearized quantum gravity framework is altered or abandoned.

\begin{figure}[b]
    \includegraphics[width=1\linewidth]{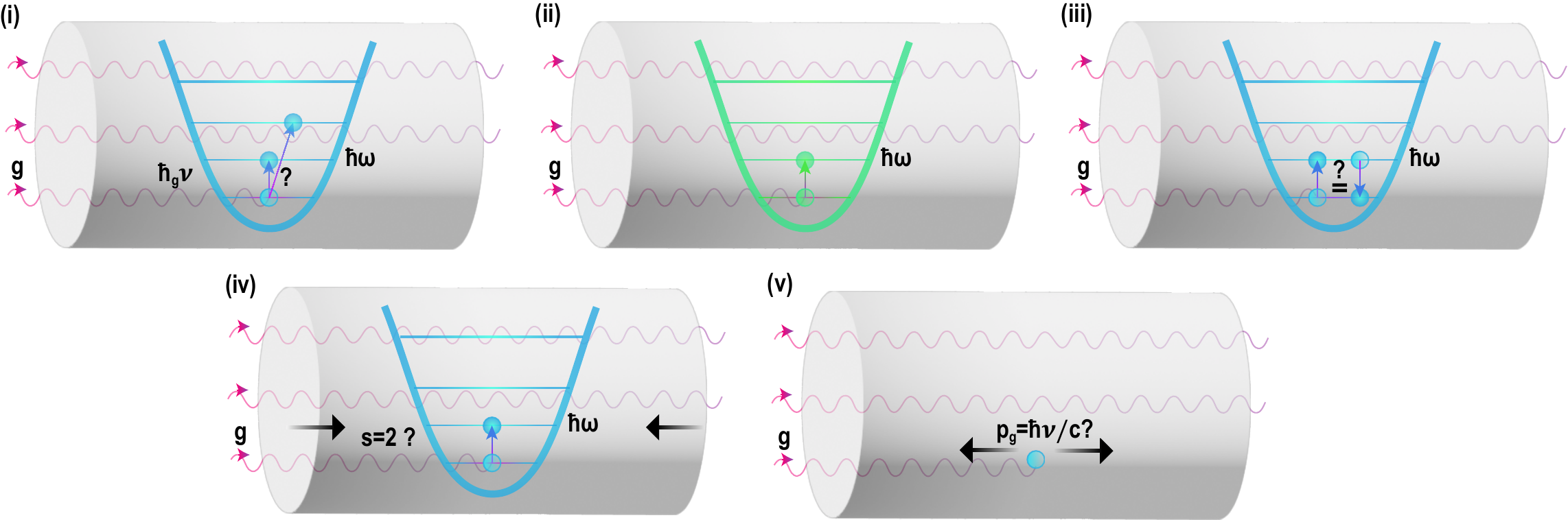}
    \caption{ 
    Detection of single gravitons from passing waves by stimulated processes can provide experimental evidence for linearized quantum gravity. The five proposed tests in the main text are illustrated here: (i) Probing the energy content in a single quantum of gravitational energy, (ii) testing universality of the interaction, (iii) comparing stimulated emission and absorption, (iv) testing the spin of a single graviton, (v) probing the quantization of momentum.}
    \label{fig:tests}
\end{figure}

Here, rather than focusing on specific speculative models, we instead present how some of the fundamental principles of linearized quantum gravity become accessible in graviton detection experiments, directly at the level of quantum theory. With some inspiration from the first experiments on quantum theory in the early 20th century,  we motivate the following five tests of the quantum interaction between matter and gravitons (see also Fig. \ref{fig:tests}), that are enabled by the setup based on stimulated exchange of quanta between GWs and matter:

\begin{itemize}
\item[(i)] Is the energy content in the graviton the same as in a photon? In other words, is Planck's constant that governs the exchange of energy with gravity the same as for all other interactions, $\hbar_G = \hbar$? One can for example imagine a natural number $k$ such that $\hbar_G = k \hbar$, and thus the energy exchange with matter proceeds as $E = k h f$. One would then expect not the transition to the first excited state, but to the k-th excited state. Similarly, for a general $\hbar_G \neq \hbar$, one can monitor for seemingly off-resonant excitations for which the frequencies between wave and matter would match to $f_G = \frac{\hbar}{\hbar_G} f$. We note that this would not violate any current experimental data on classical gravitational wave detection, and neither from other non-gravitational parts of quantum physics, as the standard constant $\hbar$ is unaltered in the non-gravitational sector. The reasoning for probing whether $\hbar_G = \hbar$ is to test whether the individual quanta of gravity are of the same nature as the quanta of light and matter.
\item[(ii)] Is $\hbar_G$ universal, or does it depend on the properties of matter and gravitational waves? This would replicate some of the early tests of quantum theory, measuring $\hbar$ for various materials. It is conceivable that the quantized interaction with gravity would follow different rules than other interactions. Such non-universality for the gravitational interaction would of course also imply (i) above, since $\hbar$ is known to be universal.
\item[(iii)] Is the probability of stimulated absorption the same as the probability of stimulated emission? This would amount to testing the gravitational Einstein coefficients, and whether $B_{12}^{(G)} = B_{21}^{(G)}$. Such a test would require the preparation of the resonator not just in the ground state, but also in the first excited state, monitoring quantum jumps to the ground state due to stimulated emission of gravitons. Since both stimulated emission and absorption are both accessible with the type of detectors we propose, this test would provide a window into the nature of energy conservation for the exchange of individual quanta.
\item[(iv)] Does the absorption of gravitons follow the quadrupole-formula as expected? Making use of independent LIGO detections, one can test if the probability of single graviton events is indeed as expected from a quadrupolar interaction with the collective mode, as given by equation \eqref{eq:prob}. This would provide experimental input on possible scalar or vector components of the interaction at the quantum level, or otherwise confirm the spin-2 nature of the particle.
\item[(v)] Do gravitational waves also carry quantized momentum $p = h f/c$? This, for example, is central to quantum noise from gravitons in interferometers \cite{parikh2021signatures}, but such fluctuations remain very difficult to observe. However, if such observations of graviton noise or imprints of gravitons on the CMB \cite{krauss2014using} could be made, in combination with tests of the $E=h f$ relation as we proposed, they would firmly establish a particle picture as they test complementary aspects of quantization, analogously to photons \cite{pais1979einstein, compton1923quantum}. Together such tests could therefore provide a bona-fide test of quantum gravity.
\end{itemize}

This list is by no means exhaustive. For example, one can also consider tests of the wave-particle duality through statistical properties, as evident in Einstein's early works \cite{einstein1906theorie, einstein1909gegenwartigen, einstein1917quantentheorie}. These examples show the potential of a single-graviton-detector as we proposed for fundamental explorations of the interaction between matter and gravity at the single graviton level. They can highlight some of the very fundamental properties that any broader theory of quantum gravity would have to obey, offering empirical anchoring of some of the basic principles. Importantly, these tests require only access to stimulated processes, and are thus attainable once single graviton absorption can be achieved. The additional requirements are the ability to detect various energy levels and modes in the resonator, prepare different detector materials and energy eigenstates as initial states, and to be able to correlate to classical detections via LIGO or other means. Thus, while such experiments with stimulated processes cannot serve as definite proofs, they clearly provide a wide range of opportunities to test aspects of quantum gravity. While we do not anticipate deviations from expected physics, these tests would provide empirical experimental evidence where none exists so far, and there is always room for surprises (sometimes leading to unexpected explanations for other open problems in physics). Furthermore, together with tests of gravitational shot noise from fluctuations in interferometers or the CMB, they could rule out most semi-classical models which otherwise might explain each effect in isolation, and thus can provide very strong evidence of quantization. 

\section{Conclusions} 

We conclude with a positive outlook on experimental tests of quantum gravity. Similar to the development of the early quantum theory of the 20th century, the mid 21st century will likely see first experimental demonstrations of the quantized interaction between gravity and matter, such as through our proposed single graviton exchange \cite{tobar2023detecting}. Other experimental proposals exist as well that probe other aspects of quantum features, such as testing how quantum superpositions source Newtonian gravity and cause entanglement \cite{PhysRevLett.119.240401,marletto2017gravitationally,krisnanda2020observable,pedernales2022enhancing}, or testing speculative phenomenological models \cite{marshall2003towards,bekenstein2012tabletop,pikovski2012probing,bassi2013models}. Here we have argued that a single graviton detector would provide a complementary approach and offer a range of opportunities to test the very basic principles of quantized interaction between gravitational waves and matter, in close analogy to historic experiments that established the existence of the photon in the early 20th century. Combined with other tests, the quantum nature of gravity can be probed from a variety of perspectives. We are thus hopeful that there are now realistic paths for experimental input to quantum gravity from laboratory experiments. 

\acknowledgements
We thank Frank Wilczek for discussions and Isabella Marotta for her assistance in reviewing historic works on the photoelectric effect. This work was supported by the National Science Foundation under grant 2239498.  G. T. acknowledges support from the General Sir John Monash Foundation. S.K.M. was supported by the Wallenberg Initiative on Networks and Quantum Information (WINQ).


%

\pagebreak
\appendix
\section{Detector response at finite duration} \label{sec:response}

In classical gravitational wave literature  \cite{MTW, rothman2006gravitons, maggiore2007}, a resonant bar detector will be modeled with damping
\begin{equation}
    \ddot{\xi}(t) + \gamma \dot{\xi}(t) + \omega^2 \xi(t) = F_G(t)/M_{eff}
\end{equation}
where $\xi(t)$ represents the mass' perturbation from equilibrium position, $\omega$ the resonance of the detector, $\gamma = \omega/Q$ the damping as a factor of the quality factor $Q$, $M_{eff}$ the effective mass of the mode, and $F_G(t)$ the effective gravitational tidal force due to the incident gravitational wave.
For $F_G(t) = -2L h \nu^2/\pi^2 \ \textrm{Re}\left[ e^{-i \nu t} \right]$, this produces the solution \cite{maggiore2007}:
\begin{equation}
 \xi(t) = \, \frac{2Lh \nu^2}{\pi^2} \frac{\left(\nu^2 - \omega^2 \right) \cos(\nu t) - \gamma \nu \sin(\nu t)}{\left(\nu^2 - \omega^2 \right)^2 + \gamma^2 \nu^2} \, .
\end{equation}
This result might seem problematic for the resonant $\nu = \omega$, non-damping $\gamma = 0$ case, however there is a specific solution for the exact resonance      without damping \cite{maggiore2007}
\begin{equation} \label{eq:mono-solution} 
\xi_{res}(t) = \frac{2 L h \omega t}{\pi^2} \sin(\omega t) \, .
\end{equation}
A perfect detector will keep absorbing energy from the gravitational wave ad infinitum, or in practice, until the damping of the detector becomes a significant factor as the system reaches a semi-stable state \cite{maggiore2007, MTW}.

However, for the binary mergers detected by LIGO, the duration of the GW in any particular frequency range is very short, and chirps upwards in frequency until the merger is finalized \cite{abbott2016observation,abbott2017gw170817}. The incident gravitational wave persists in the resonant band of our detector only for some short time $\tau$, and thus the solution \eqref{eq:mono-solution} is valid for time $\tau$ without damping playing the limiting role. Applying this time scale to the resonant result, we can calculate the expected graviton excitation probability using the relation $\frac{dP}{dt} = \frac{1}{\hbar \omega} \frac{dE}{dt}$ \cite{rothman2006gravitons}:
\begin{equation}
P_{0 \rightarrow 1} \simeq \frac{E}{\hbar \omega} = \frac{M_{eff}}{2 \hbar \omega} \Bigl( \left\langle \dot{\xi}(t)^2 \right\rangle + \omega^2 \left\langle \xi(t)^2 \right\rangle \Bigr) 
= \frac{h^2 M L^2 \omega^3 \tau^2}{\hbar \pi^4}
\end{equation}
where $M_{eff} = M/2$ for a simple bar. 
This result agrees with the quantum excitation probability computed in Ref. \cite{tobar2023detecting} and given in eq. \eqref{eq:prob} within the finite duration window $\tau$.

\end{document}